\documentclass[12pt]{article}
\usepackage{amsmath,amsfonts,epsf}
\usepackage{amssymb}
\usepackage{graphicx}
\usepackage{grffile}
\input epsf

\def\be{\begin{equation}}
\def\ee{\end{equation}}
\def\bea{\begin{eqnarray}}
\def\eea{\end{eqnarray}}
\def\ie{\begin{equation}\begin{aligned}}
\def\fe{\end{aligned}\end{equation}}

\renewcommand{\title}[1]{\vbox{\center\LARGE{#1}}\vspace{5mm}}
\renewcommand{\author}[1]{\vbox{\center#1}\vspace{5mm}}
\newcommand{\address}[1]{\vbox{\center\em#1}}
\newcommand{\email}[1]{\vbox{\center\tt#1}\vspace{5mm}}

\begin{document}
\begin{titlepage}
\begin{center}
\hfill \\
\hfill \\
\vskip 1cm

\title{A Duality Appetizer}

\author{Daniel Jafferis $^{a}$ and
Xi Yin$^{b}$}

\address{Center for the Fundamental Laws of Nature
\\
Jefferson Physical Laboratory, Harvard University,\\
Cambridge, MA 02138 USA}

\email{$^a$jafferis@physics.harvard.edu,
$^b$xiyin@fas.harvard.edu}

\end{center}

\abstract{ We propose that the three-dimensional ${\cal N}=2$ $SU(2)$ Chern-Simons theory at level 1 coupled to an adjoint chiral multiplet with no superpotential is equivalent to the free field theory consisting of a single massless ${\cal N}=2$ chiral multiplet. In particular, we show that the two theories have the identical ``$Z$-function" and identical superconformal index.
}

\vfill

\end{titlepage}

Remarkable progress has been made recently on strongly coupled three dimensional superconformal field theories using supersymmetric localization \cite{Pestun, Kapustin:2009kz, Kapustin:2010xq, Drukker, Jafferis:Z, Kim:2009wb, Imamura, Bashkirov:2011pt}. In this note we present a computation of the $Z$-function \cite{Jafferis:Z} and superconformal index \cite{Imamura} of ${\cal N}=2$ $SU(2)$ Chern-Simons theory at level $k=1$, coupled to one adjoint chiral matter multiplet $\Phi$.\footnote{Theories of this type have been previously studied in \cite{Gaiotto:2007qi, Niarchos:2008jb, Niarchos:2009aa}. } Due to miraculous identities, we will find that they coincide with that of a single (ungauged) free massless ${\cal N}=2$ chiral multiplet $X$. In particular, by minimizing the $Z$-function we find that the R-charge of $\Phi$ is renormalized to $\Delta=1/4$ at $k=1$, and therefore ${\rm Tr}\Phi^2$ has dimension $1/2$ and is a free field. This leads us to propose that the ${\cal N}=2$ $SU(2)_1$ Chern-Simons theory with one adjoint matter {\sl is} the free theory of $X\simeq {\rm Tr}\Phi^2$, plus possibly a topological sector. 

To begin, let us consider the $SU(2)$ Chern-Simons theory at level $k$ coupled to $M$ adjoint matter fields, with no superpotential. The $Z$-function, which is defined as the partition function of the theory on $S^3$ with a choice of R-charge $\Delta$, is computed by the matrix model (which reduces to a one-dimensional integral for $SU(2)$)
\ie
Z_{k,M}(\Delta) =  \int du \sinh^2(2\pi u) e^{2 \pi i k u^2} e^{M\left[ \ell(1-\Delta)+\ell(1-\Delta+2 i u) + \ell(1 - \Delta - 2 i u) \right]} 
\fe
where $\ell(z)$ is the double sine function, given by
\ie
& \ell(z) = -z \log(1-e^{2\pi i z}) + {i\over 2} \left[ \pi z^2 + {1\over \pi} {\rm Li}_2(e^{2\pi iz}) \right] - {\pi i\over 12},\\
& \ell'(z) = -\pi z \cot(\pi z).
\fe
While we do not know a general method to evaluate this integral analytically, it can be easy performed numerically by rotating the integration contour to $e^{i\theta} {\mathbb R}$, with some nonzero phase $\theta$ such that the integrand falls off exponentially at the two ends of the contour. The superconformal R-charge of the IR CFT is then given by the value of $\Delta$ that locally minimizes $|Z_{k,M}(\Delta)|^2$. We denote this value by $\Delta_{k,M}$. For $1\leq k\leq 4$ and $1\leq M\leq 4$, we find the following numerical results for $\Delta_{k,M}$:\footnote{$\Delta$ takes the value $1/2$ in the free limit $k\to \infty$. Our result in the $M=1$ case is consistent with the bound of \cite{Niarchos:2009aa} which was argued by considering superpotential deformations and the $s$-rule. }
\begin{center} \begin{tabular}{| l | c | c | c  | r | } \hline  & M=1 & M=2 & M=3 & M=4 \\ \hline k=1 & .25  & .303587 & .359644 &  .394141 \\ \hline k=2 & .291765  & .333243 & .371037  & .398601 \\ \hline k=3 &  .326463 & .355965 &  .38259  & .404209 \\ \hline k=4 & .35593 & .374585 & .393374  & .410183 \\ \hline \end{tabular} \end{center}
In particular, at $k=M=1$, the renormalized R-charge of $\Phi$ is $1/4$. While we have not proven this result analytically, it is confirmed numerically to high accuracy. It then follows that the gauge invariant operator ${\rm Tr}\Phi^2$ has dimension $1/2$ which saturates the unitarity bound, and hence must be a free scalar field, provided that the ${\cal N}=2$ IR fixed point exists and there are no accidental global symmetries.

Even more surprisingly and unexpectedly, the entire $Z$-function $Z_{1,1}(\Delta)$ matches that of a free chiral multiplet up to a simple phase factor, thanks to the following identity
\ie
\int du \sinh^2(2\pi u) e^{2 \pi i u^2} e^{\ell(1-\Delta)+\ell(1-\Delta+2 i u) + \ell(1 - \Delta - 2 i u)} = {1\over 2\sqrt{2}} e^{{\pi i\over 2}(1+\Delta)^2-{\pi i\over 4}}  e^{\ell(1-2 \Delta)}
\fe
We found this identity numerically and confirmed it to high accuracy. A proof should be possible along the lines of \cite{Teschner06}. In other words, the relation between the $Z$-function of the $SU(2)_1$ CS + 1 adjoint theory and the free chiral multiplet is
\ie
Z_{1,1}(\Delta) = {1\over 2\sqrt{2}} e^{{\pi i\over 2}(1+\Delta)^2-{\pi i\over 4}} Z_X(\Delta).
\fe
The simple phase factor difference between the $Z$-functions of the two theories could be due to the contribution from an additional topological sector, e.g. a decoupled abelian Chern-Simons.  The sphere partition function of abelian Yang-Mills-Chern-Simons at level 2 is given by 
\ie
Z(\Delta_B) = \int dv\, e^{2 \pi i v^2} e^{2\pi \Delta_B v} = {e^{{\pi i\over 4}}\over \sqrt{2}} e^{{\pi i\over 2}\Delta_B^2},
\fe
where the R-charge is given by $R = R_0 + \Delta_B B$ in terms of the classical R-charge and the topologial (monopole) charge, B.  We propose that $\Delta_B$ should be identified with $\Delta+1$. 
It is perhaps unsurprising that such a topological sector is needed since the $SU(2)_1$ Chern-Simons-matter theory violates parity while the theory of free chiral multiplet $X$ is parity invariant.

To further check the proposed equivalence, we shall compute the superconformal index \cite{Witten, Sen, Romelsberger:2005eg, Kinney}
\ie
I(x,z) = {\rm Tr}(-)^F e^{-\beta (H-R-j_3)} x^{H+j_3} z^{\tilde F}
\fe
where $F$ is the fermion number, $\beta$ is a positive real parameter, $H$ is the scaling dimension, $R$ the superconformal $U(1)$ R-charge, $j_3$ the third component of the spin, and $\tilde F$ the global $U(1)$ flavor symmetry with the assignment of charge $1/2$ to $\Phi$. The trace is taken over the spectrum of operators, or equivalently, the Hilbert space of states on $S^2$. It is a consequence of supersymmetry that the index is independent of $\beta$. Supersymmetric localization gives the following result for the theory with one adjoint flavor and CS level $k$:  \cite{Imamura}
\ie\label{indf}
I(x,z) = \sum_{s\in{\mathbb Z}} \int_0^{2\pi} {da\over 4\pi} e^{-ikas} x^{-2\Delta |s|} z^{-2|s|} \exp\left[ 
\sum_{m=1}^\infty {1\over m} f(e^{ima},x^m,z^m;s) \right]
\fe
where $\Delta$ is the renormalized superconformal R-charge of $\Phi$, and
\ie
& f(e^{ia},x,z;s) = f_\Phi(e^{ia},x,z;s) + f_V(e^{ia},x;s),\\
& f_V(e^{ia},x;s) = -e^{ia} x^{2|s|} - e^{-ia} x^{-2|s|},\\
& f_\Phi(e^{ia},x,z;s) = z{x^\Delta\over 1-x^2} \left( 1+e^{ia} x^{2|s|} + e^{-ia} x^{2|s|} \right)
- z^{-1} {x^{2-\Delta}\over 1-x^2} \left( 1+e^{ia} x^{2|s|} + e^{-ia} x^{2|s|} \right).
\fe
In (\ref{indf}) the exponentials involving $f_\Phi$ and $f_V$ are the 1-loop determinants from chiral multiplet $\Phi$ and the vector multiplet, respectively. The integer $s$ labels the magnetic flux sectors on the $S^2$. In the case $k=1$, with $\Delta=1/4$, we find the result as a power series in $x$:
\ie
I_{k=1}(x,z) &= 1+z x^{1\over 2} + z^2 x + (-z^{-1}+z^3) x^{3\over 2} + (-1+z^4)x^2 + z^5 x^{5\over 2}
+ z^6 x^3 \\
&~~~~+ (-z^{-1}+z^7) x^{7\over 2} + (-2+z^8) x^4 + {\cal O}(x^{9\over 2})
\fe
which is in complete agreement with the superconformal index for a free chiral multiplet $X$ with the assignment of flavor charge $\tilde F=1$, namely,
\ie
I_X(x,z) = \exp\left[ 
\sum_{m=1}^\infty {1\over m} f_0(x^m,z^m) \right],~~~~f_0(x,z) = z{x^{1\over 2}\over 1-x^2} - z^{-1} {x^{3\over 2}\over 1-x^2}.
\fe
Furthermore, we find that only the $s=0$ and $s=\pm 1$ sectors contribute to $I_{k=1}(x,z)$. Note that $s$ and $-s$ are gauge equivalent and were simply counted twice in the sum in (\ref{indf}). Their respective contributions are
\ie
&I^{s=0}_{k=1} (x,z) =1+z x^{1\over 2} + z^2 x + z^3 x^{3\over 2} + (-1+z^4)x^2 + z^5 x^{5\over 2}
+ z^6 x^3 \\
&~~~~~~~ + z^7 x^{7\over 2} + z^{-{1\over 2}} x^{15\over 4} + (-2+z^8) x^4 + {\cal O}(x^{9\over 2}) ,
\\
&I^{s=\pm 1}_{k=1} (x,z) = -z^{-1} x^{3\over 2} - z^{-1} x^{7\over 2} - z^{-{1\over 2}} x^{15\over 4} +  {\cal O}(x^5).
\fe
Note a few peculiar features of these contributions. There is a term $z^{-{1\over 2}} x^{15\over 4}$ from the $s=0$ sector contributions that cancels against a similar term of the opposite sign in the $s=\pm1$ sector contribution. There are no states of such R charge or scaling dimension in the free chiral multiplet theory. If we deform the Chern-Simons-matter theory by adding Yang-Mills coupling so that it is free in the UV, such states would be present in the UV theory. Presumably, they are lifted in the IR CFT, which must be the case if our conjecture is correct.

Also, remarkably, the terms $zx^{1\over 2}-z^{-1}x^{3\over 2}$, which comes from the single letter operators $X$ and $(\bar\psi_X)_+$ (the subscript $+$ labels the $j_3=1/2$ component of the fermion), comes from two distinct magnetic flux sectors in the $SU(2)_1$ CS-matter theory, namely $s=0$ and $s=\pm 1$ ! This is consistent because in the $SU(2)_1$ theory there is no gauge invariant topological charge associated with the magnetic flux.

Let us comment on a brane construction. The ${\cal N}=2$ $U(N_c)_k$ CS theory with one adjoint can be embedded in type IIB string theory as the low energy theory on two D3-branes stretched between an NS5-brane and a $(1,k)$ 5-brane \cite{Kitao:1998mf, Bergman:1999na} (see also \cite{Kutasov}). The NS5-brane is extended along the directions 012456, while the $(1,k)$ 5-brane is extended along 01245 and at an angle in the 6-9 plane so as to preserve ${\cal N}=2$ supersymmetry. They are separated along $x^3$. The D3-branes extend in 0123 directions, and are free to move in the 4-5 plane. Since all fields are in the adjoint, the overall $U(1)$ decouples. One may attempt to move the NS5-brane across the $(1,k)$ 5-brane \cite{Jafferis-Yin}, but this is subtle because the 5-branes are not transverse to one another. One may deform the configuration by rotating say the $(1,k)$-brane in the 4-7, 5-8 and 6-9 planes in such a way that the ${\cal N}=2$ supersymmetry is preserved and a mass term is generated for the adjoint chiral multiplet. The $s$-rule \cite{Hanany-Witten} would then indicate that supersymmetry is spontaneously broken for $k< N_c$. In particular, $N_c=2, k=1$ belong to this case. If we deform the $SU(2)_1$ CS-matter theory by a superpotential mass term ${\rm Tr}\Phi^2$, it should correspond to deforming the superpotential by $W=X$ in the dual free theory. The latter indeed leads to $F$-term supersymmetry breaking.

Let us also comment that the $SU(2)$ CS-matter theory at level $k=2,3$ and $M=1$ has renormalized R-charge $\Delta<1/3$. The allows a relevant superpotential deformation by ${\rm Tr}\Phi^6$ (in addition to the ${\rm Tr}\Phi^4$ deformation which is relevant perturbatively), which leads to new critical points.

To conclude, we found highly nontrivial evidence for the conjecture that the ${\cal N}=2$ $SU(2)_1$ Chern-Simons theory with one adjoint matter is equivalent to a free field theory of a single ${\cal N}=2$ chiral multiplet. Such a duality is reminiscent to those of \cite{Seiberg, Aharony:1997bx, deBoer:1997kr, Kutasov, Niarchos:2008jb, Jafferis-Yin}. It would be nice to prove analytically the identities involving $Z$-functions we found. While we have demonstrated that this CS-matter CFT, if exists, must contain a free chiral multiplet $X\simeq {\rm Tr}\Phi^2$, there is still in principle the possibility of a remaining isolated strongly interacting CFT with trivial chiral ring, trivial $Z$-function and trivial superconformal index. 
In previously known examples/conjectures of IR dualities of strongly coupled theories, the matching of moduli spaces of vacua in the infrared is a main piece of evidence. In our case we can determine the geometry of the moduli space (i.e. it is smooth at the origin) using supersymmetric localization method, which would be otherwise difficult since it is not determined by ${\cal N}=2$ supersymmetry.
If our conjecture is correct, it is likely that this duality is a special case of a more general family of equivalences involving strongly coupled Chern-Simons-matter superconformal field theories. 

\bigskip

Acknowledgments: We are grateful to Shiraz Minwalla for inspiring discussions on a related problem. This work is supported in part by the Fundamental Laws Initiative Fund at Harvard University.
X.Y. is supported in part by NSF Award PHY-0847457.


\begin{thebibliography}{}


\bibitem{Pestun} 
V.~Pestun,
``Localization of gauge theory on a four-sphere and supersymmetric Wilson loops,"
arXiv:0712.2824 [hep-th].

\bibitem{Kapustin:2009kz}
A.~Kapustin, B.~Willett and I.~Yaakov,
``Exact Results for Wilson Loops in Superconformal Chern-Simons Theories with Matter,"
JHEP {\bf 1003}, 089 (2010)
[arXiv:0909.4559 [hep-th].

\bibitem{Kapustin:2010xq}
  A.~Kapustin, B.~Willett and I.~Yaakov,
  ``Nonperturbative Tests of Three-Dimensional Dualities,''
  JHEP {\bf 1010}, 013 (2010)
  [arXiv:1003.5694 [hep-th]].

\bibitem{Drukker}
N.~Drukker, M.~Marino and P.~Putrov,
``From weak to strong coupling in ABJM theory,"
arXiv:1007.3837 [hep-th].

\bibitem{Jafferis:Z}
D.~L.~Jafferis,
``The Exact Superconformal R-Symmetry Extremizes Z,"
arXiv:1012.3210 [hep-th].

\bibitem{Kim:2009wb}
  S.~Kim,
  ``The Complete superconformal index for N=6 Chern-Simons theory,''
  Nucl.\ Phys.\  {\bf B821}, 241-284 (2009).
  [arXiv:0903.4172 [hep-th]].

\bibitem{Imamura}
Y.~Imamura, S.~Yokoyama,
``Index for three dimensional superconformal field theories with general R-charge assignments,"
arXiv:1101.0557 [hep-th].

\bibitem{Bashkirov:2011pt}
D.~Bashkirov and A.~Kapustin,
``New and old N=8 superconformal field theories in three dimensions,"
arXiv:1103.3548 [hep-th].

\bibitem{Gaiotto:2007qi}
  D.~Gaiotto and X.~Yin,
  ``Notes on superconformal Chern-Simons-Matter theories,''
  JHEP {\bf 0708}, 056 (2007)
  [arXiv:0704.3740 [hep-th]].
  
\bibitem{Niarchos:2008jb}
  V.~Niarchos,
  ``Seiberg Duality in Chern-Simons Theories with Fundamental and Adjoint
  Matter,''
  JHEP {\bf 0811}, 001 (2008)
  [arXiv:0808.2771 [hep-th]].
  
\bibitem{Niarchos:2009aa}
  V.~Niarchos,
  ``R-charges, Chiral Rings and RG Flows in Supersymmetric Chern-Simons-Matter
  Theories,''
  JHEP {\bf 0905}, 054 (2009)
  [arXiv:0903.0435 [hep-th]].

\bibitem{Teschner06}
A.~G.~Bytsko and J.~Teschner,
``Quantization of models with non-comapct quantum group symmetry: Modular XXZ magnet and lattice sinh-Gordon model,"
J.\ Phys.\ A {\bf 39}, 12927 (2006)
[arXiv:hep-th/0602093].

\bibitem{Witten}
  E.~Witten,
  ``Dynamical Breaking Of Supersymmetry,''
  Nucl.\ Phys.\  B {\bf 188}, 513 (1981).

\bibitem{Sen}
  D.~Sen,
  ``Witten Index Of Supersymmetric Chiral Theories,''
  Phys.\ Rev.\  D {\bf 39}, 1795 (1989).

\bibitem{Romelsberger:2005eg}
  C.~Romelsberger,
  ``Counting chiral primaries in N = 1, d=4 superconformal field theories,''
  Nucl.\ Phys.\  {\bf B747}, 329-353 (2006).
  [hep-th/0510060].

\bibitem{Kinney}
  J.~Kinney, J.~M.~Maldacena, S.~Minwalla and S.~Raju,
  ``An index for 4 dimensional super conformal theories,''
  Commun.\ Math.\ Phys.\  {\bf 275}, 209 (2007)
  [arXiv:hep-th/0510251].

\bibitem{Hanany-Witten}
  A.~Hanany and E.~Witten,
  ``Type IIB superstrings, BPS monopoles, and three-dimensional gauge
  dynamics,''
  Nucl.\ Phys.\  B {\bf 492}, 152 (1997)
  [arXiv:hep-th/9611230].
  
\bibitem{Kitao:1998mf}
  T.~Kitao, K.~Ohta and N.~Ohta,
  ``Three-dimensional gauge dynamics from brane configurations with
  (p,q)-fivebrane,''
  Nucl.\ Phys.\  B {\bf 539}, 79 (1999)
  [arXiv:hep-th/9808111].
  
\bibitem{Bergman:1999na}
  O.~Bergman, A.~Hanany, A.~Karch and B.~Kol,
  ``Branes and supersymmetry breaking in 3D gauge theories,''
  JHEP {\bf 9910}, 036 (1999)
  [arXiv:hep-th/9908075].

\bibitem{Kutasov}
  A.~Giveon and D.~Kutasov,
  ``Seiberg Duality in Chern-Simons Theory,''
  Nucl.\ Phys.\  B {\bf 812}, 1 (2009)
  [arXiv:0808.0360 [hep-th]].

\bibitem{Seiberg}
K.~A.~Intriligator, N.~Seiberg,
``Mirror symmetry in three-dimensional gauge theories,"
Phys.\ Lett.\ {\bf B387}, 513-519 (1996)
[hep-th/9607207].

\bibitem{Aharony:1997bx}
  O.~Aharony, A.~Hanany, K.~A.~Intriligator, N.~Seiberg, M.~J.~Strassler,
  ``Aspects of N=2 supersymmetric gauge theories in three-dimensions,''
  Nucl.\ Phys.\  {\bf B499}, 67-99 (1997).
  [hep-th/9703110].
  
\bibitem{deBoer:1997kr}
  J.~de Boer, K.~Hori and Y.~Oz,
  ``Dynamics of N = 2 supersymmetric gauge theories in three dimensions,''
  Nucl.\ Phys.\  B {\bf 500}, 163 (1997)
  [arXiv:hep-th/9703100].

\bibitem{Jafferis-Yin}
D.~L.~Jafferis and X.~Yin,
``Chern-Simons-Matter Theory and Mirror Symmetry,"
arXiv:0810.1243 [hep-th].




\end{thebibliography}
\end{document}